\documentclass[namedreferences]{solarphysics}
\usepackage[optionalrh]{spr-sola-addons}
\usepackage{epsfig}
\usepackage{graphicx}
\usepackage{color}
\usepackage{url}
\usepackage{lscape}
\usepackage{sidecap}
\newcommand{\etal}{{\it et al.}}
\usepackage[pdfborder={0 0 0 },urlcolor=blue]{hyperref}
\ifx \doiurl \undefined \def \doiurl#1{\href{http://dx.doi.org/#1}{\url{#1}}}\fi
\ifx \adsurl \undefined \def \adsurl#1{\href{http://adsabs.harvard.edu/abs/#1}{\url{#1}}}\fi

\begin{document}
\begin{article}
\begin{opening}

\title{Long-term Periodicities in  North--South Asymmetry of Solar Activity and
Alignments of the Giant Planets}

\author{J. Javaraiah}

\institute{\#58, Bikasipura (BDA Layout), Bengaluru-560111,  India.\\
Formerly working at Indian Institute of Astrophysics, Bengaluru-560 034, India.\\
email: \url{jajj55@yahoo.co.in;  jdotjavaraiah@gmail.com}\\
}

\runningauthor{J. Javaraiah}
\runningtitle{Periodicities in North--south Asymmetry of Solar Activity}

\begin{abstract}
The existence of $\approx$12-year and $\approx$51-year periodicities
 in the north--south asymmetry of solar activity is well known.  
 However, the origin of these
 as well as  the well-known relatively short 
 periodicities in  the north--south asymmetry is not yet clear.
Here we have analyzed the combined daily data of sunspot groups reported in  
Greenwich Photoheliographic Results (GPR) and Debrecen Photoheligraphic Data
 (DPD) during the period 1874\,--\,2017 and the data of 
the orbital positions (ecliptic longitudes) of the giant planets in 
  ten-day intervals during the period 1600\,--\,2099. 
Our analysis suggests that $\approx$12-year and $\approx$51-year periodicities
in the north--south asymmetry of solar activity are the manifestations 
 of the  differences in the strengths of $\approx$11-year
and $\approx$51-year periodicities of activity in the northern- and
 southern-hemispheres. During the period 1874\,--\,2017 
the Morlet wavelet power spectrum of the north--south asymmetry of 
 sunspot-group area and that of the
 mean absolute difference  ($\overline{\psi_{\rm D}}$)  of the
orbital positions of the giant planets are found to be similar. 
Particularly, there is a suggestion that  
 the  $\approx$12-year and  $\approx$51-year periodicities 
 in the north--south asymmetry of sunspot-group area   occurred
 during approximately the same times as the corresponding periodicities
 in $\overline{\psi_{\rm D}}$. 
 Therefore, we
suggest that there could be influence of some specific configurations of 
the giant planets in  the origin of the 
 $\approx$12-year and $\approx$50-year periodicities of the north--south 
asymmetry of solar activity.  
\end{abstract}


\end{opening}

\section{Introduction}
The existence of the $\approx$12-year and 40\,--\,50-year  periodicities
 in the  north--south asymmetry of solar activity has been reported by a number
 of authors.\linebreak  \inlinecite{cob93}  analyzed the combined  
 GPR (1874\,--\,1983) and  United States Air Force (1983\,--\,1989) 
data of sunspot groups and found that   only a 12.1-year periodicity is 
 statistically significant in the north--south asymmetry of sunspot area.
  \inlinecite{yi92} was first to point out that in the
north--south asymmetry, derived by using the conventional 
formula  $\frac{N - S}{N + S}$ (relative asymmetry),    
 the 11\,--\,12-year periodicity is mostly  an artifact of the 11-year
 cycle of the denominator $N + S$, 
where $N$ and $S$ are the quantities  of the activity  in 
northern and southern hemispheres, respectively.   
\inlinecite{jg97}  verified this by 
determining separately the power spectra of the $\frac{N - S}{N+S}$
and the $N - S$ (absolute asymmetry) of the sunspot data. They  
 found that a peak at $\approx$$\frac{1}{12}$ year$^{-1}$ frequency  
 in the power spectrum of $N-S$ is statistically insignificant and 
  the same peak in the spectrum of 
$\frac{N-S}{N+S}$ is
 statistically very significant 
(see Figure~7 in \opencite{jg97}). Obviously, they  
have  confirmed the doubt of 
\inlinecite{yi92}. 
\inlinecite{boc05} have  done the same.
However, the statistical error in the relative asymmetry, $\frac{N - S}{N+S}$,
  is much
smaller than  that
of the absolute asymmetry, $N-S$ (\opencite{jg97}). Therefore, 
variations of the relative asymmetry could be  more reliable than those 
 of the absolute asymmetry.
 \inlinecite{jg97} have also  shown the existence of  
 $\approx$50-year  periodicity in both  the relative asymmetry 
  and the  absolute asymmetry, and a 8\,--\,9-year periodicity in the 
absolute asymmetry. Similar periods were also noticed by
\inlinecite{boc05}. \inlinecite{knaack04} by  using a wavelet analysis,
studied the long-term periodicities in the north--south asymmetry  
of the  monthly averaged sunspot areas (1874\,--\,2003)  and found the 
existence of the $\approx$12-year and  $\approx$43-year periodicities in 
the relative asymmetry of the sunspot area.
Similar periodicities in the relative north--south asymmetry 
of the areas of different size sunspot groups were also seen by
 \inlinecite{mb16}, using  the Kodaikanal white-light digitized
 archive sunspot observations (1921\,--\,2011). \inlinecite{deng16} from
 the wavelet analysis of sunspot areas of Solar Cycles 
9\,--\,14  found the existence of $\approx$9-year 
and $\approx$51-year periodicities in both  the relative asymmetry and
the absolute asymmetry of sunspot area, and a $\approx$12-year periodicity 
in  the relative asymmetry only. 
 Recently, \inlinecite{jj19} studied  cycle-to-cycle modulations in the
relative north--south asymmetry of solar-cycle maximum and minimum and
 found that 
there exists a 30\,--\,50-year periodicity in the 
asymmetry of minimum and a possibility of the existence of a much longer 
 (mostly substantially  more than 100 years) periodicity in 
the asymmetry of maximum. Overall,   the existence of 
8\,--\,12-year and 40\,--\,50-year periodicities in the north--south 
asymmetry of solar activity is somewhat established  from 
 observational studies. 
\inlinecite{verma93} detected the existence of a  110-year 
periodicity in the relative north--south asymmetry of solar activity (also see 
 \opencite{li02}; \opencite{li19}). 
\inlinecite{bo11} by using the method of Spectral Variation Analysis 
 found the existence of long 
periods of phase shifts between northern- and southern-hemisphere's activity. 

 The physical reason  of the long periodicities
 as well as  the well-known relatively short 
 periodicities (170\,--\,180-day, 320\,--\,329-day, 1.5-year, 1.8-year,
  2.1-year,  3.6-year, $etc.$; see \opencite{knaack04}; \opencite{rj15} 
and the references therein)  in  the 
north--south asymmetry is not yet clear.
 Recently,  
\inlinecite{sc18} showed that  around 8.5-year and 30\,--\,50-year 
 periods in the  absolute asymmetry are  the sum and beat periods of
 a 22-year period  magnetic  dipolar mode  and a 13\,--\,15-year period 
 quadrupolar mode.
 \inlinecite{nep19} showed that   long-term variations in the
 north--south asymmetry of solar activity 
 resulting from short-term fluctuations in the $\alpha$-effect of solar 
dynamo.
On the other hand, a number of authors suggested the existence of  a connection  between  solar variability (both long- and short-terms)
and planetary configurations~(\opencite{jose65}; \opencite{wood65};
 \opencite{wood72}; \opencite{gj95}; \opencite{zaq97}; \opencite{juc03}; 
\opencite{wolff10}; \opencite{abreu12}; \opencite{cc12}; \opencite{wil13};
\opencite{salv13}; \opencite{sharp13}; \opencite{chow16}; \opencite{stef16};
 \opencite{stef19}).
\inlinecite{jj05} found  the existence of a good
agreement between
the  amplitudes of the variations in the Sun's spin and the orbital
angular momenta at the common epochs of the steep decreases
in  both the orbital angular momentum and the Sun's equatorial
rotation rate determined from sunspot data.
Since the Sun's equator is 
inclined $7.17^\circ$ to the ecliptic, the influences of 
  configurations of the planets on solar activity may be different 
across the Sun's equator (also see \opencite{juc00}).  
Here we have analyzed the combined daily data of sunspot groups reported in  
 GPR and DPD during the period 1874\,--\,2017 and
the orbital positions (ecliptic longitudes) of the giant planets
available for each ten-day interval during the period 1600\,--\,2099 
and found that during the period 1874\,--\,2017 
the Morlet wavelet power spectrum of the north--south asymmetry of 
sunspot-group area and that of the mean absolute difference  
($\overline{\psi_{\rm D}}$)  
of the orbital positions of the giant planets are  similar.
The wavelet spectra
 suggest that  12\,--\,13-year and the 40\,--\,50-year periodicities  in 
the north--south asymmetry of  sunspot-group area exist 
during approximately the same times as the corresponding periodicities 
 in $\overline{\psi_{\rm D}}$.
 Therefore, we
suggest that there could be  influence of some configurations of 
the giant planets in  the origin of 
 $\approx$12-year and 40\,--\,50-year periodicities of the north--south 
asymmetry of solar activity.  
 
In the next section we describe the data analysis, in Section~3 we present
the results,  and in Section~4 we summarize the conclusions and discuss
 them briefly.

\section{Data analysis}
Recently, \inlinecite{jj19} analyzed the  GPR and DPD sunspot-group
daily data  during the period April 1874\,--\, June 2017 
(downloaded from {\sf fenyi.solarobs.unideb.\break hu/pub/DPD/}) and 
 derived the time series of the 13-month smoothed monthly
mean corrected whole-spot areas of the sunspot groups in the Sun's whole sphere
(WSGA), northern hemisphere (NSGA), and southern hemisphere (SSGA). 
Here we have used  these time series and derived the 
corresponding time series of both the relative  north--south asymmetry (RNSA) 
and absolute  north--south asymmetry (ANSA). 
The time series of the 13-month smoothed  monthly mean values
of the international sunspot number  $R_{\rm Z}$ (ISSN)
during the period 1874\,--\,2014 are  downloaded  from
{\sf www.sidc.be/silso/datafiles}
(Source: WDC-SILSO, Royal Observatory of Belgium, Brussels).
Here we have  used the data on ecliptic  longitudes (in degrees)
$\psi_{\rm J}$, $\psi_{\rm S}$, $\psi_{\rm U}$, and $\psi_{\rm N}$
of the giant planets Jupiter (J), Saturn (S), Uranus (U), and Neptune (N)
in each 10-day interval during the period 1600\,--\,2099.
These data were provided by  Ferenc Varadi.
He had derived these data using the  Jet
Propulsion Laboratory (JPL) DE405
ephemeris~(\opencite{seid92}; \opencite{stan98})
for the period 1600\,--\,2099.
We determined  the average value ($\overline{\psi_{\rm D}}$)
of the absolute differences
$|\psi_{\rm J}-\psi_{\rm S}|$, $|\psi_{\rm J}-\psi_{\rm U}|$, 
 $|\psi_{\rm J}-\psi_{\rm N}|$, $|\psi_{\rm S}-\psi_{\rm U}|$,
 $|\psi_{\rm S}-\psi_{\rm N}|$, and $|\psi_{\rm U}-\psi_{\rm N}|$
in each ten-day interval during the period 1600\,--\,2099. 
The values of the  absolute differences that are greater than $180^\circ$ are 
converted as $360^\circ$ minus the absolute difference. This is  because of 
here  we  use minimum unsigned  difference  between the ecliptic
longitudes (have values $0^\circ$\,--\,$360^\circ$) of any two planets   
(we have not done this in the preliminary
results reported in \opencite{jj18}, where we compared the
variations in the amplitudes of solar cycles and  $\overline{\psi_{\rm D}}$). 
From the wavelet transformation both time and frequency localization can be 
obtained (for more detail see \opencite{tc98}). 
 We have used the IDL codes of  Morlet wavelet  
 power spectral analysis, cross-wavelet transformation (XWT), and wavelet 
coherence (WCOH) provided  by \inlinecite{tc98}.
We have  downloaded these  from {\tt paos.colorado.edu/research/wavelets}.
The cross-wavelet  spectrum 
reveals localized
 similarity (covariance) in time and scale (period).
WCOH is a normalized time and scale (period) resolved 
measure for the relationship between two time 
series (also see \opencite{mk04}).  A number of authors
 used XWT and WCO  to study asynchronous behavior of northern and southern 
hemispheres activity 
($e.g.$, \opencite{li10}; \opencite{deng16}, and the references therein).
We made  Morlet wavelet power spectral analysis of NSGA, SSGA, RNSA, and
ANSA and compared the power spectra. We also made the Morlet wavelet spectral 
analysis of  $\overline{\psi_{\rm D}}$ and  compared   the wavelet power spectra of 
$\overline{\psi_{\rm D}}$, RNSA, and ANSA.  In addition, we have applied
 XWT and WCOH to the time series of these parameters to confirm the 
similarities seen in the Morlet wavelet power spectra of these parameters. 

\begin{figure}
\centering
\includegraphics[width=\textwidth]{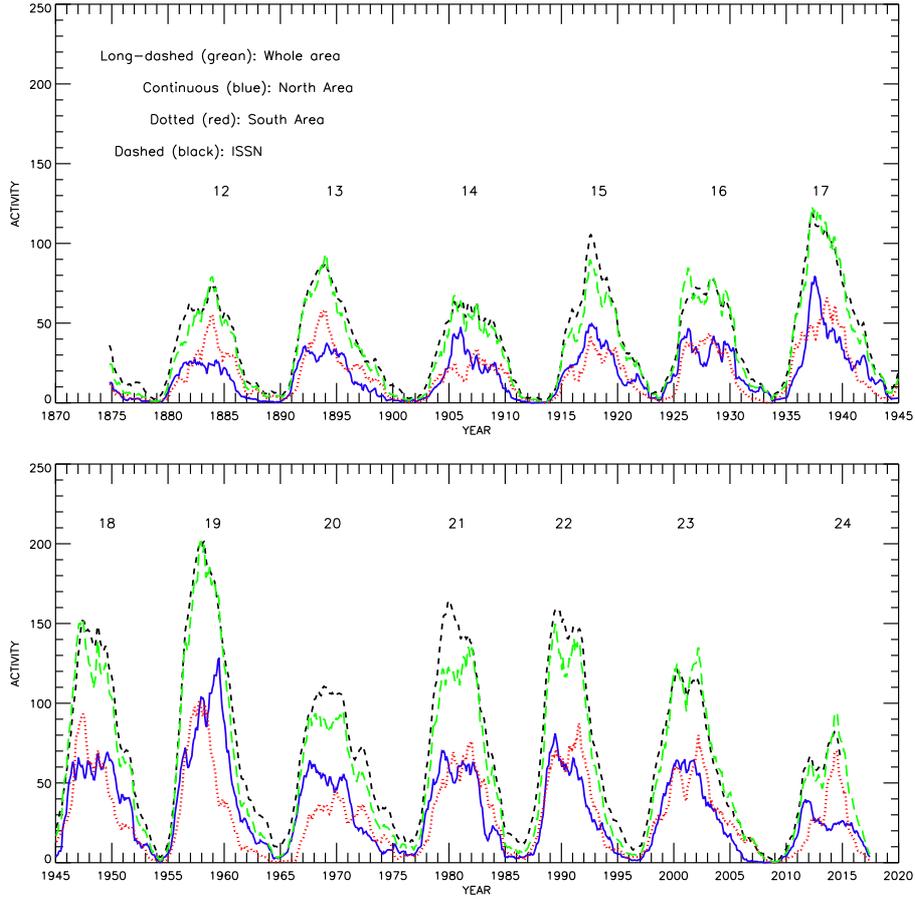}
\caption{Variations in the 13-month smoothed monthly mean area 
 of sunspot groups in the whole 
sphere: WSGA ({\it green-long-dashed curve}),
 northern hemisphere: NSGA ({\it blue-continuous curve}), 
southern hemisphere: SSGA ({\it red-dotted curve}) during the period 
1874\,--\,2017, and international
sunspot number (ISSN) $R_{\rm_Z}$ ({\it black-dashed curve}) during the period
1874\,--\,2014. 
The values of WSGA, NSGA, and SSGA
 are first divided by the largest value of WSGA, 3480.15 msh, and then
multiplied by
the largest value, 201.3, of ISSN. Waldmeier numbers of the solar cycles
are also shown. This figure is
a slightly  modified (clarity improved) version of  the Figure~1 of
Javaraiah (2019).}

\label{f1}
\end{figure}

\begin{figure}
\centering
\includegraphics[width=\textwidth]{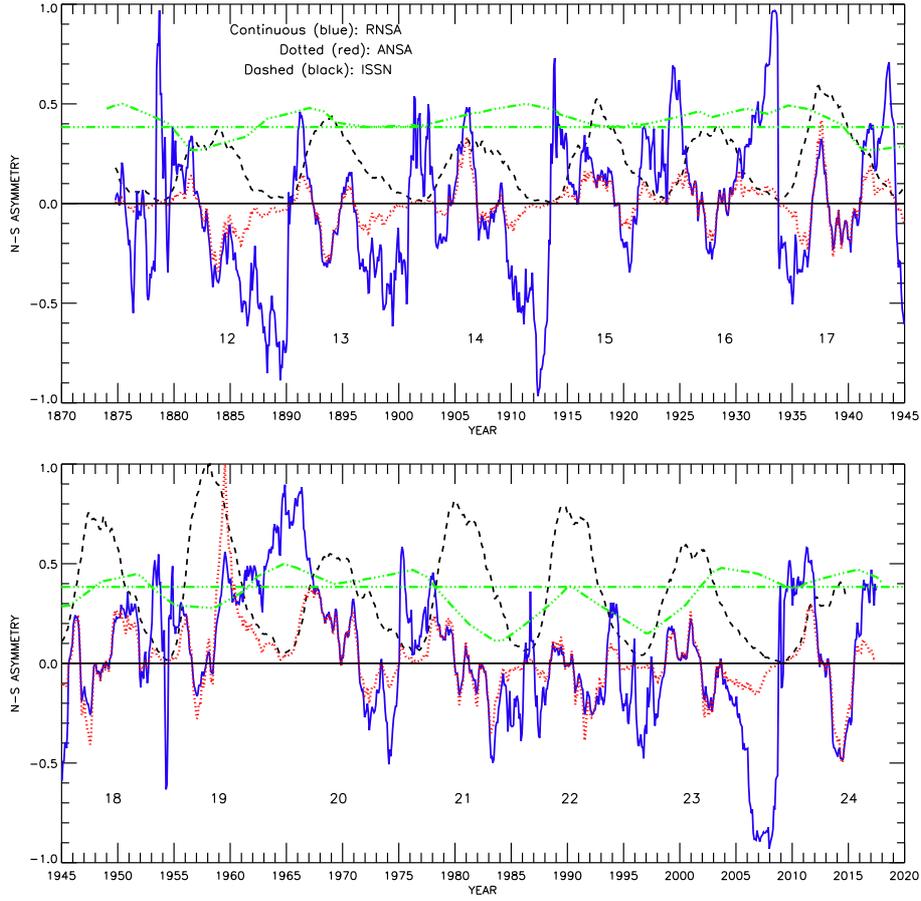}
\caption{Variations in the relative north--south asymmetry: RNSA 
({\it blue-continuous curve}) and  absolute north--south asymmetry:
 ANSA ({\it red-dotted curve}) 
 determined from  13-month smoothed monthly NSGA and SSGA during the period 
1874\,--\,2017.
The latter is divided by the maximum value (1593.61 msh) of the absolute ANSA. 
A positive (negative) value implies  that the activity  in northern (southern)
 hemisphere is higher than  that of in southern (northern) hemisphere.  
 The {\it black-dshed curve} represents the variation in the normalized
 international sunspot number (ISSN) $R_{\rm_Z}$ during the period 
1874\,--\,2014 and 
the {\it green-dotted-dashed curve} represent the  $\overline{\psi_{\rm D}}$ in
 ten-day intervals during the period 1874\,--\,2017 (divided by two times of
maximum value, $119.05^\circ$). Waldmeier numbers of the solar cycles 
are also shown.}
\label{f2}
\end{figure}

\begin{figure}
\centering
\includegraphics[width=12.0cm]{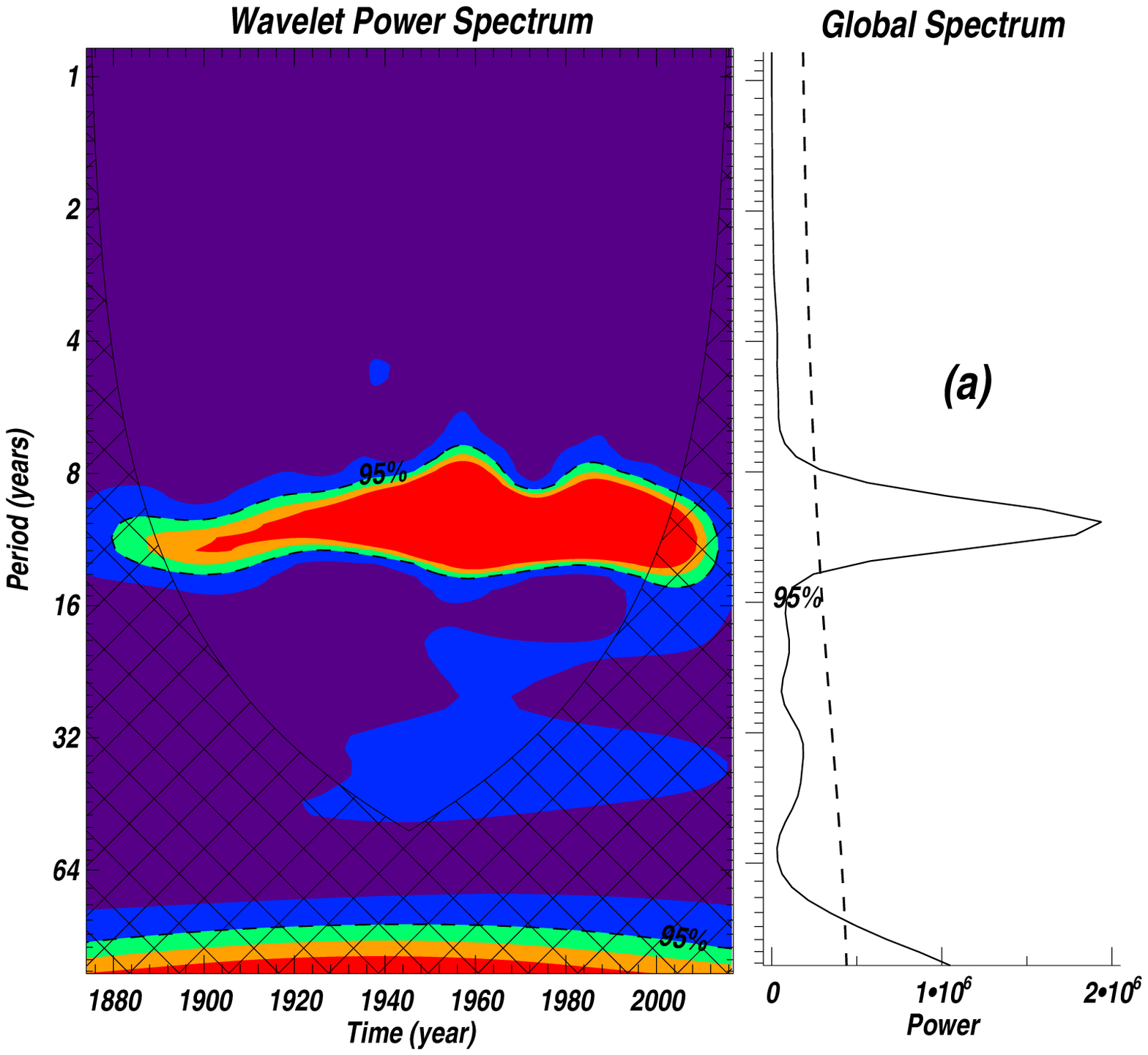}
\includegraphics[width=12.0cm]{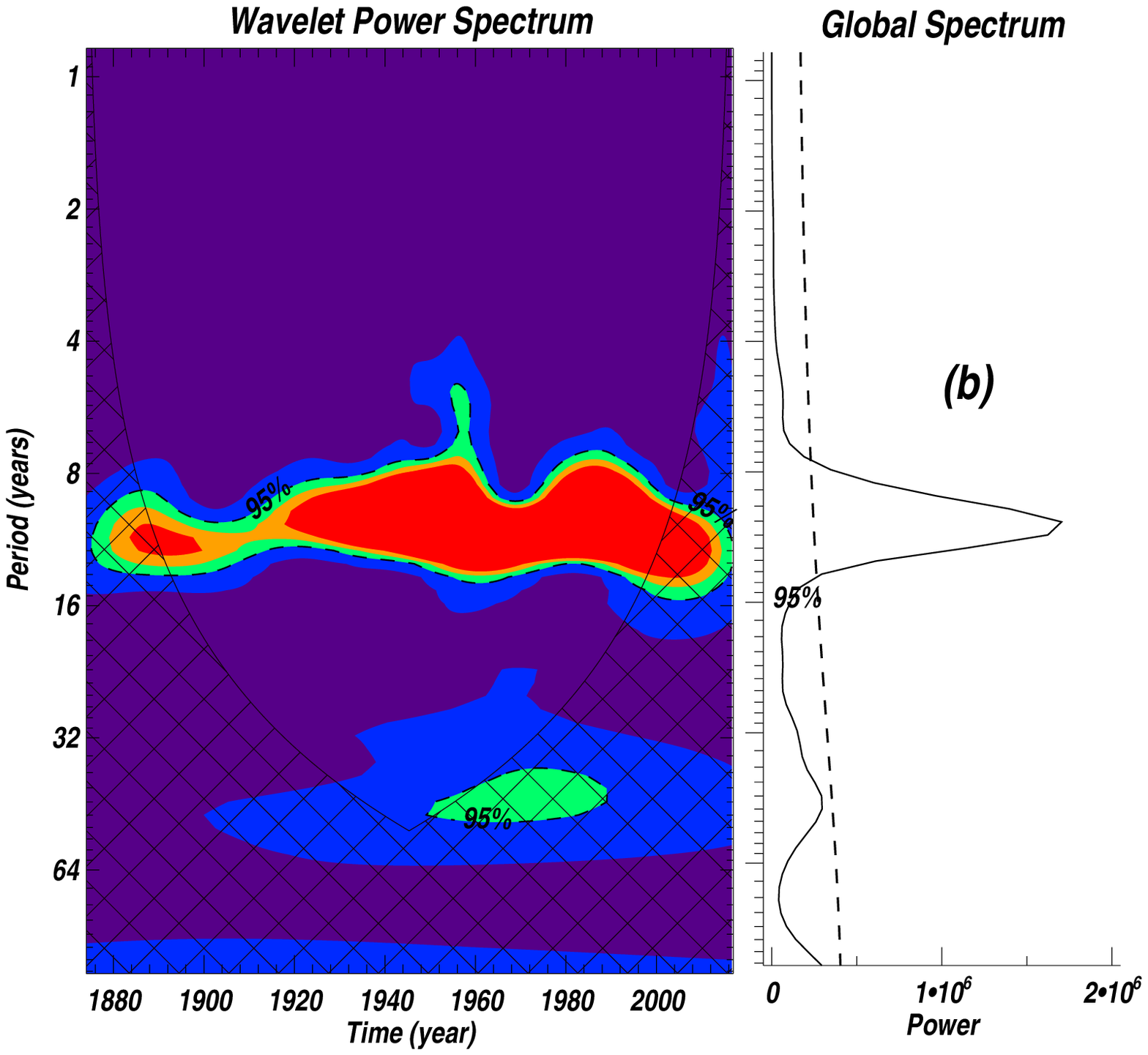}
\caption{Panels ({\bf a}) and ({\bf b}) show the wavelet and global
power spectra of  NSGA and SSGA
shown in Figure~1, respectively (power is divided by 10).
The wavelet spectra are  normalized
by the variances of the corresponding time series. The shadings are  at
the normalized variances of 1.0, 3.0, 4.5, and 6.0.
The {\it dashed curves} represent the 95\% confidence levels
deduced by assuming a white-noise process.
The {\it cross-hatched regions} indicate the cone of
influence where edge effects become significant (Torrence and Compo, 1998).}
\label{f3}
\end{figure}

\begin{figure}
\centering
\includegraphics[width=12.0cm]{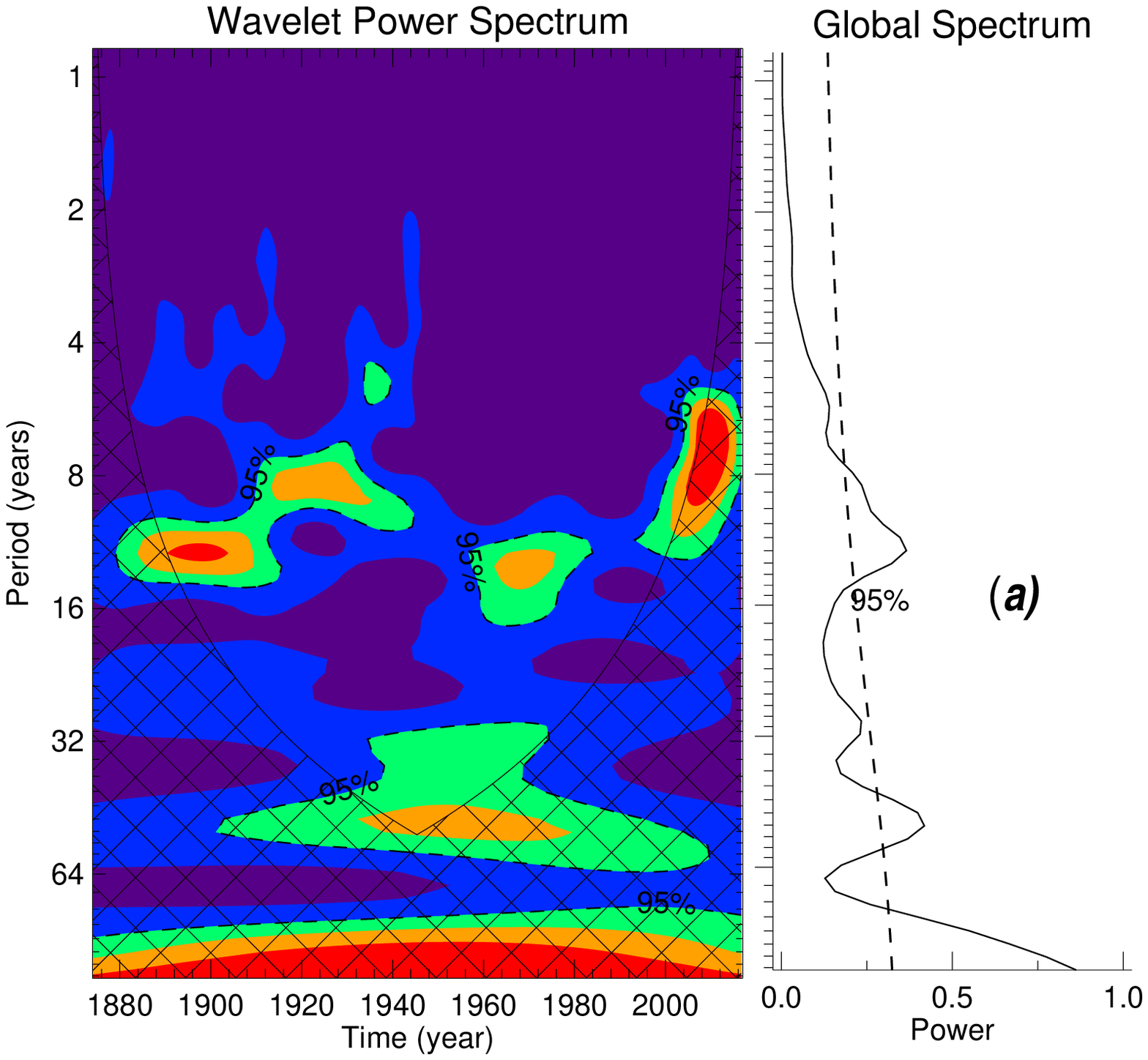}
\includegraphics[width=12.0cm]{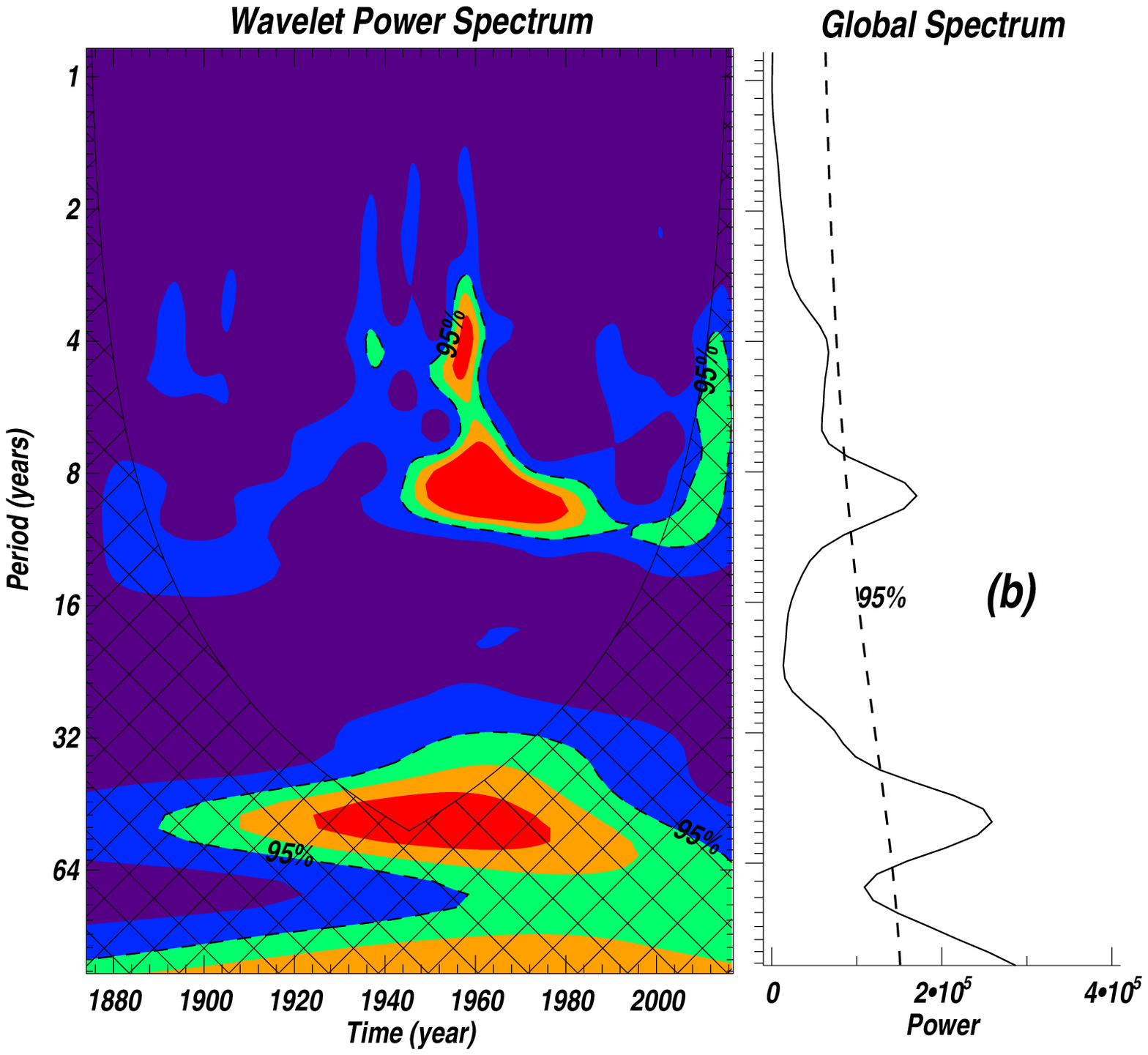}
\caption{Panels ({\bf a}) and ({\bf b}) show the wavelet and global
power spectra of RNSA  and ANSA
shown in Figure~2, respectively (power is divided by 10).
The wavelet spectra are  normalized
by the variances of the corresponding time series. The shadings are  at
the normalized variances of 1.0, 3.0, 4.5, and 6.0.
The {\it dashed curves} represent the 95\% confidence levels
deduced by assuming a white-noise process.
The {\it cross-hatched regions} indicate the cone of
influence where edge effects become significant (Torrence and Compo, 1998).}
\label{f4}
\end{figure}

\begin{figure}
\centering
\includegraphics[width=12.0cm]{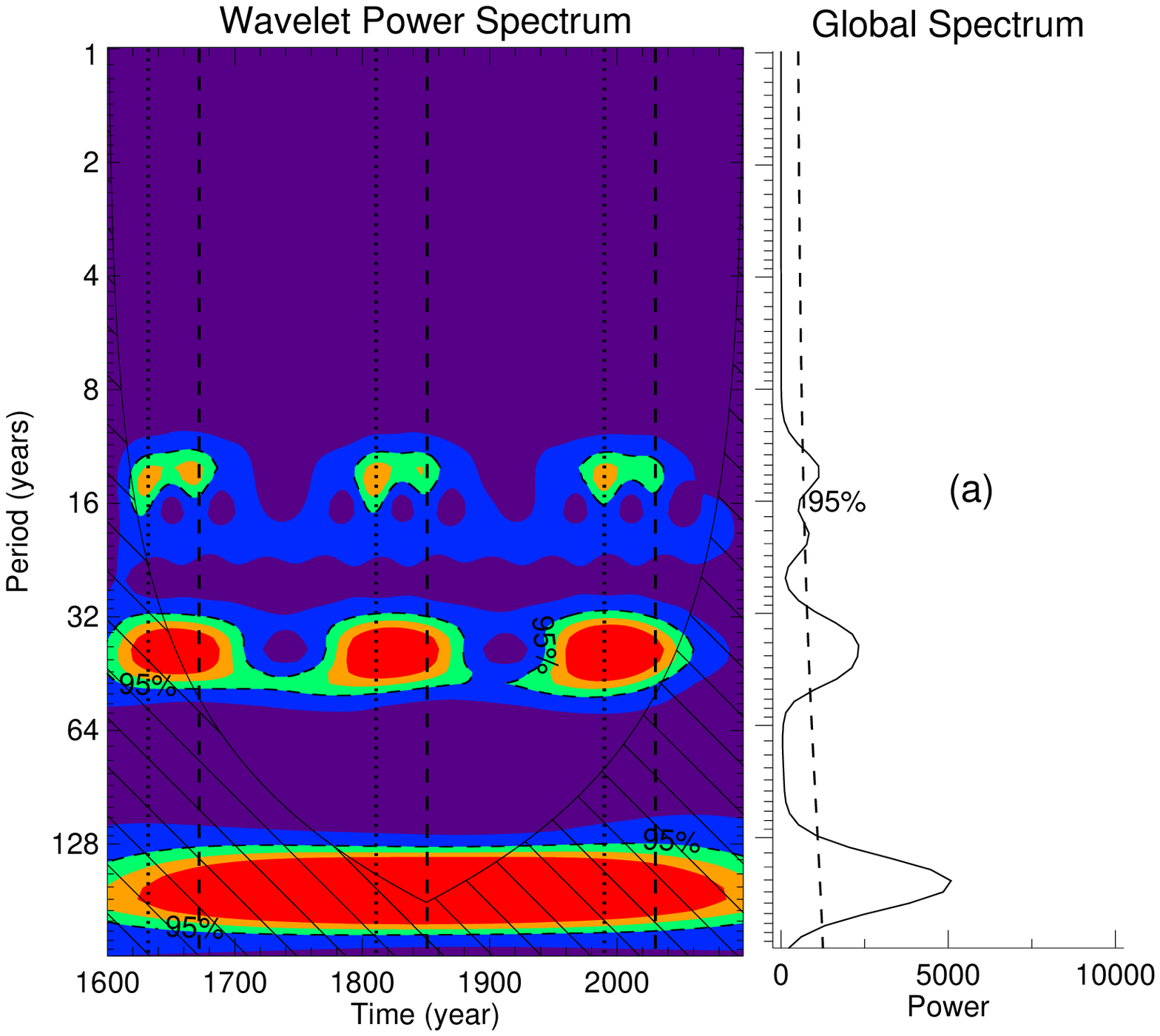}
\includegraphics[width=12.0cm]{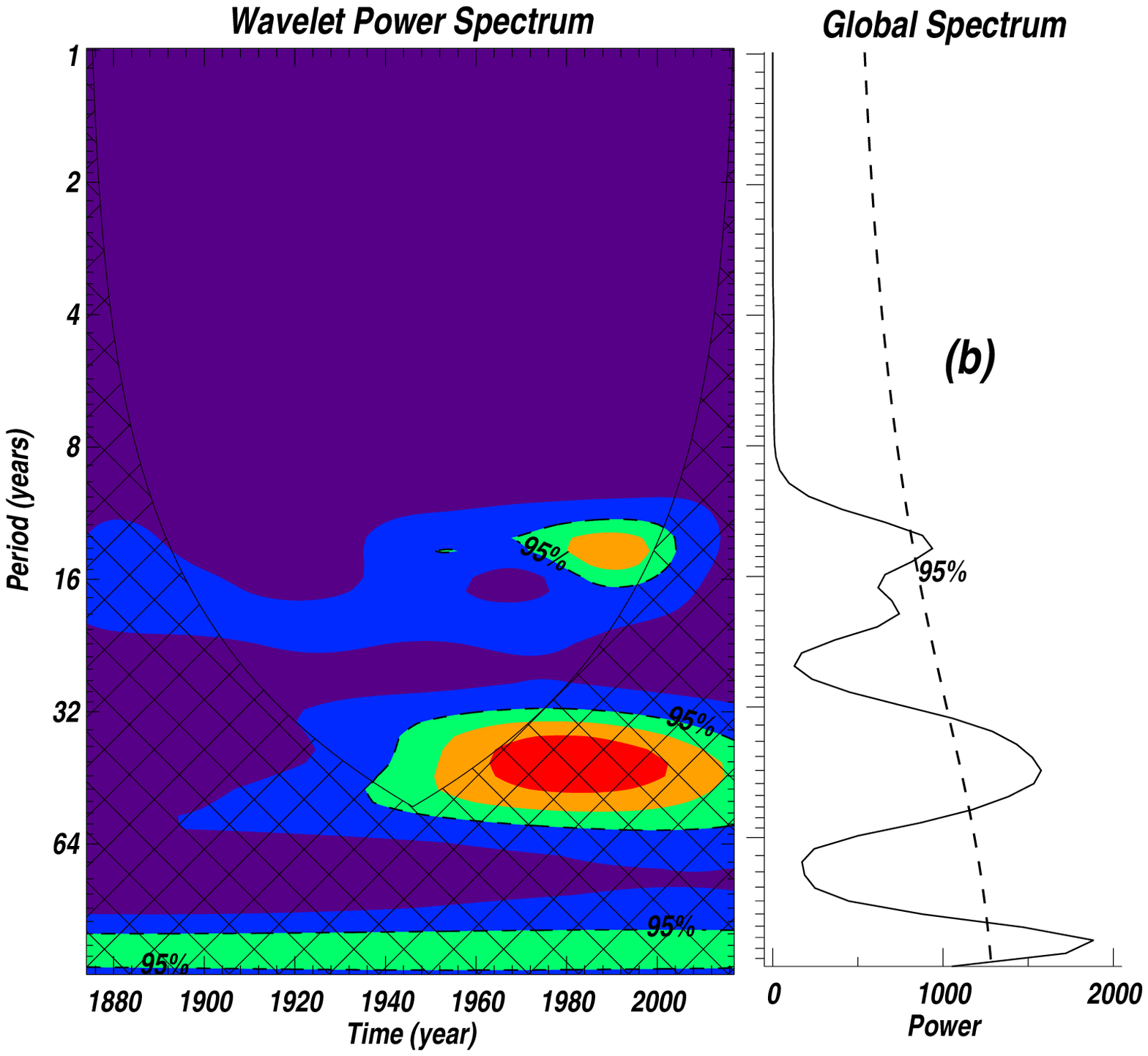}
\caption{Panels ({\bf a}) and ({\bf b}) show the wavelet and global
power spectra of   $\overline{\psi_{\rm D}}$ in ten-day intervals during 
the periods 1600\,--\,2099 and 1874\,--\,2017 (that is shown in Figure~2), 
respectively (power is divided by 100).
The wavelet spectra are  normalized
by the variances of the corresponding time series. The shadings are  at
the normalized variances of 1.0, 3.0, 4.5, and 6.0.
The {\it dashed curves} represent the 95\% confidence levels
deduced by assuming a white-noise process.
The {\it cross-hatched regions} indicate the cone of
influence where edge effects become significant (Torrence and Compo, 1998).
The {\it dotted vertical lines} (at 1632, 1811, and 1990) and the  
{\it dashed vertical lines} (at 1672, 1851, and 2030)
 are drawn in the panel {\bf a}
 at the epochs of the steep decreases in the
orbital angular momentum of the Sun.  
There are some differences in the alignments of the giant planets 
 at the locations of the {\it dotted} and the 
{\it continuous vertical lines} (see Javaraiah, 2005).}
\label{f5}
\end{figure}

\begin{figure}
\setcounter{figure}{3}
\centering
\begin{subfigure}
\includegraphics[width=5.5cm]{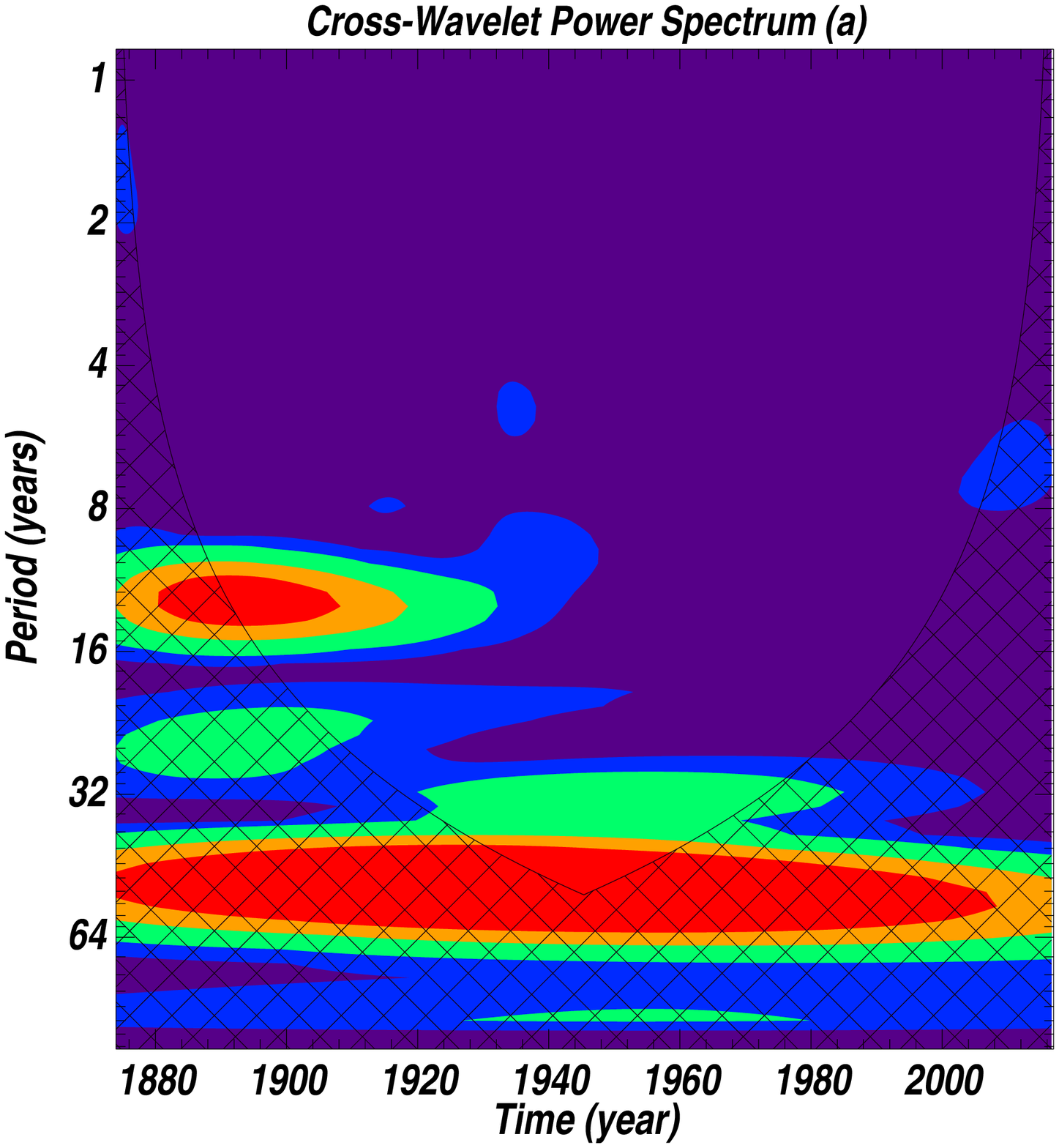}
\end{subfigure}
\begin{subfigure}
\includegraphics[width=5.5cm]{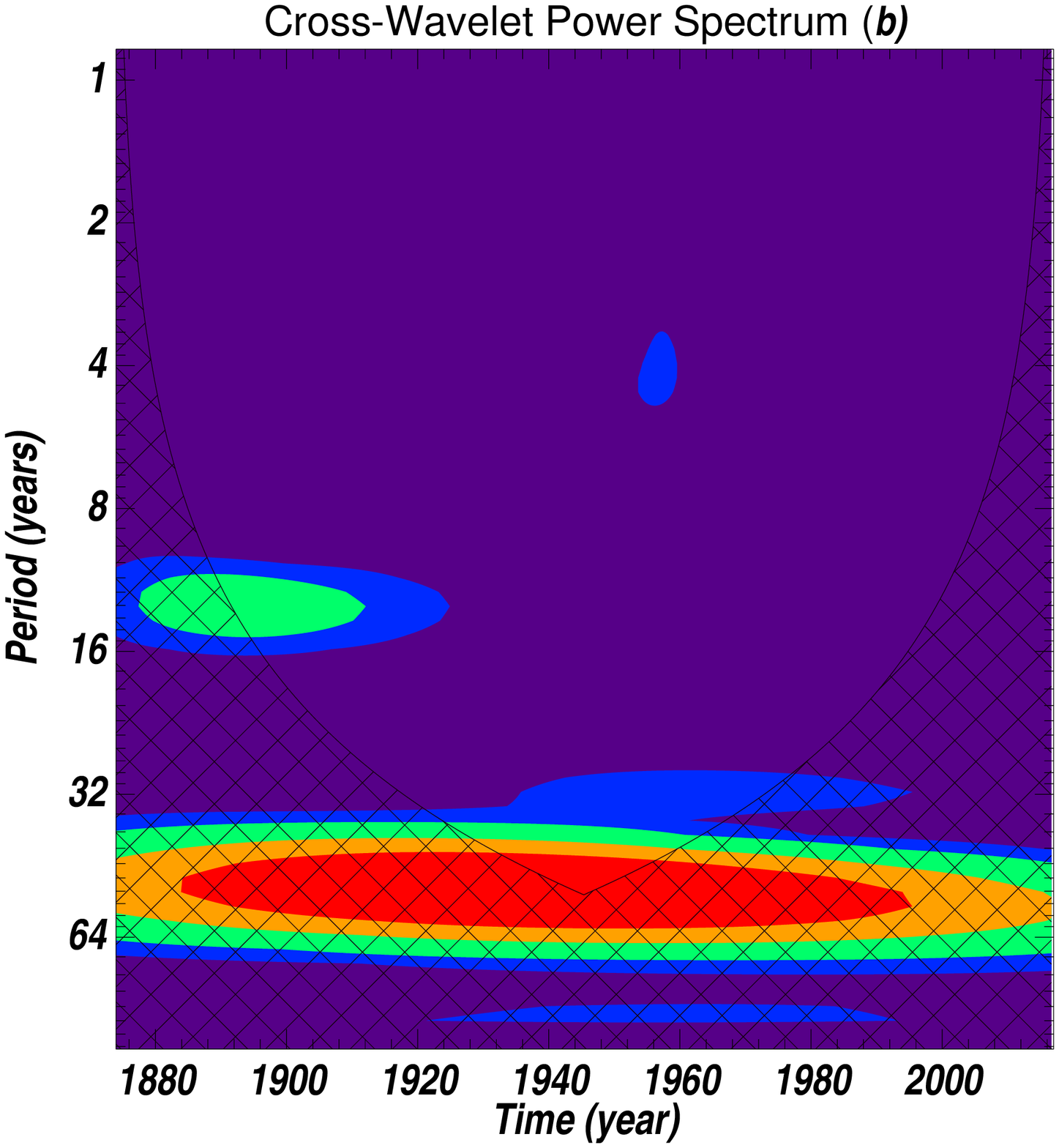}
\end{subfigure}
\caption{The panel  ({\bf a})  shows the cross-wavelet 
power spectra of RNSA and  $\overline{\psi_{\rm D}}$   and the panel 
({\bf b}) shows 
that of ANSA and $\overline{\psi_{\rm D}}$  during 
1874\,--\,2017 (in the case of panel {\bf b} power is multiplied by $10^6$).
  The shadings are  at levels
 1, 3, 10, and 20.
The {\it cross-hatched regions} indicate the cone of
influence where edge effects become significant (Torrence and Compo, 1998).}
\label{f6}
\end{figure}

\begin{figure}
\setcounter{figure}{4}
\centering
\begin{subfigure}
\includegraphics[width=5.5cm]{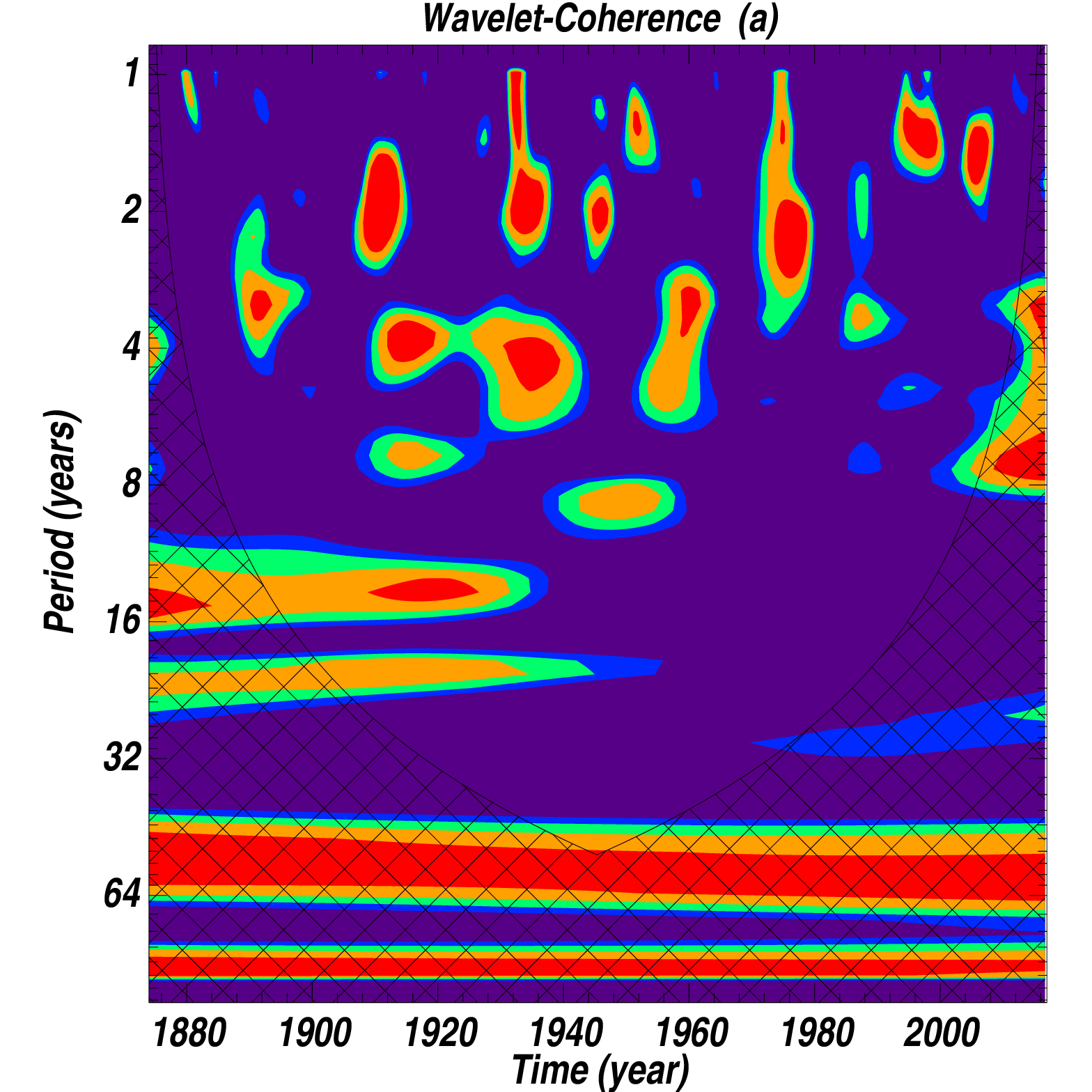}
\end{subfigure}
\begin{subfigure}
\includegraphics[width=5.5cm]{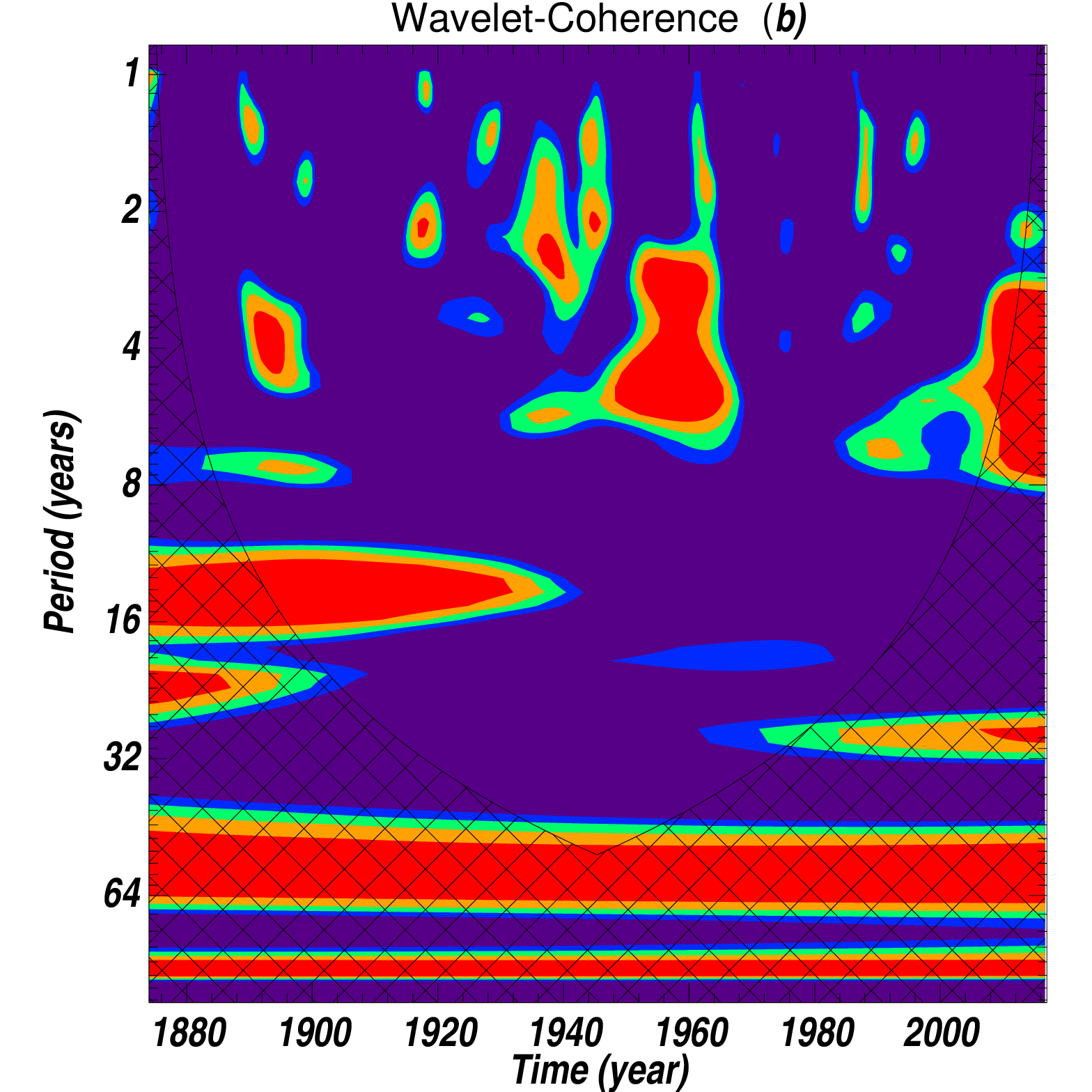}
\end{subfigure}
\caption{The panel ({\bf a}) shows the wavelet-coherence (WCOH) 
 of RNSA and  $\overline{\psi_{\rm D}}$ and the panel ({\bf b}) shows 
that of ANSA (normalized) and  $\overline{\psi_{\rm D}}$  during 
 1874\,--\,2017.  
 The shadings are  at levels 
 0.5, 0.6, 0.7, and 0.8.
A  value of WCOH  one means the existence of a linear 
relationship and zero means vanishing correlation.       
The {\it cross-hatched regions} indicate the cone of
influence where edge effects become significant (Torrence and Compo, 1998).}
\label{f7}
\end{figure}

\section{Results}
Figure~1 shows the variations in the 13-month smoothed monthly mean
 WSGA, NSGA, and SSGA during the period 1874\,--\,2017,  and the variation 
in ISSN  ($R_{\rm_Z}$) during the period 1874\,--\,2014. This figure is 
a slightly  modified (clarity improved) version of  the Figure~1 of
 \inlinecite{jj19}.
Figure~2 shows the variations in the relative north--south asymmetry, RNSA, 
 and  in the absolute north--south asymmetry, ANSA, determined from the  
 13-month smoothed monthly mean NSGA and SSGA during the period 1874\,--\,2017.
In this figure the variation in $\overline{\psi_{\rm D}}$ values of 
 ten-day intervals during the period 1874\,--\,2017 and 
 the variation in ISSN during the period 1874\,--\,2014 are also shown. 
As can be seen in this figure, both the RNSA and the ANSA are considerably
 vary. The variation in RNSA is much  more pronounced than that in ANSA. 
Obviously, the 
variations of RNSA and ANSA are 
 considerably different.  A major difference is during the minima of 
 a majority of solar cycles RNSA is much larger than that of during the maxima. 
The behavior of ANSA   is somewhat  opposite to it. However, still there 
exists a high  
 correlation between RNSA and ANSA (correlation 
coefficient $r = 0.63$).  During the late Maunder minimum
the activity pattern, $i.e.$ the large north--south asymmetry in the 
activity (see \opencite{soko94}) 
 is consistent with the aforementioned behavior of RNSA.
There are $\approx$11-year cycles in all RNSA, 
ANSA, and $\overline{\psi_{\rm D}}$. However,  
there is a suggestion that  no appreciable correlation 
between ISSN and 
any of the parameters RNSA, ANSA, and $\overline{\psi_{\rm D}}$. 
 There are patterns (around 1885\,--\,1955 and 1965\,--\,2015) of about 
40\,--\,60-year cycles in $\overline{\psi_{\rm D}}$. Such a pattern seems to 
  exist also in RNSA and ANSA, but it is not 
 visible clearly. 

Figures~3a and 3b show the Morlet wavelet power spectra, and the corresponding
global spectra, of NSGA and SSGA, respectively. As can be seen in these 
figures, obviously an $\approx$11-year periodicity  exists
 in both NSGA and SSGA and
 dominant almost throughout the period 1874\,--\,2017. There is also a 
suggestion 
on the existence of a relatively much weaker $\approx$50-year periodicity
 in both the NSGA and the SSGA, but after around 1920. 
In the Morlet 
wavelet and global spectra of WSGA, which are not shown here, besides  the 
much stronger $\approx$11-year periodicity, a very weak $\approx$50-year 
periodicity is seen.     

Figures~4a and 4b show the Morlet wavelet power spectra, and the corresponding
global spectra, of the relative north--south asymmetry, RNSA, and the absolute 
north--south asymmetry, ANSA, respectively. As can be seen in these figures
a $\approx$12.8-year periodicity exists in RNSA and a $\approx$9.6 year 
periodicity exists in ANSA. However, there is a strong suggestion that 
 each of these periods vary with time considerably (from 8 to 16 years) and 
systematically. That is, a large change of this periodicity seems to be
 occurring in alternate
30\,--\,35 years, mainly in the case of RNSA. That is, in the case of RNSA 
a $\approx$12.8-year periodicity was strong around the period 1880\,--\,1920, 
a $\approx$9-year 
period was strong around the period 1920\,--\,1950,
 a $\approx$12.8-year period was strong around the period 
1960\,--\,1990, and than onwards a $\approx$9-year periodicity  is  strong. 
Such clear alternate patterns are not present  in ANSA, but 
overall variations in  the $\approx$12.8-year periodicity of RNSA and 
in the $\approx$9-year periodicity of ANSA are the same.
In ANSA the $\approx$9-year 
periodicity was  strong only around the period 1945\,--\,1985.  
There is a suggestion on the existence of a $\approx$51-year 
periodicity (poorly resolved)  in both RNSA and ANSA and it was strongly
 present during the period 1900\,--\,1990. The global spectrum of RNSA
 suggests that the strengths of the
$\approx$12.8-year  and  the $\approx$51-year periodicities in RNSA
are almost equal (the former is only slightly weaker than the latter) in RNSA, 
whereas the global spectrum of ANSA suggests that in 
ANSA the $\approx$51-year periodicity is considerably stronger than 
the  $\approx$9-year periodicity.  
 Besides the existence of  these periodicities, 
there are episodes of a 4\,--\,5-year 
periodicity in both RNSA and ANSA. This periodicity was strong in RNSA around
1960. \inlinecite{dd96} found a strong 35-year periodicity 
in the north--south asymmetry of solar filament activity 
 during 1919\,--\,1989. 
 A  $\approx$32-year periodicity is also
present in RNSA (see Figure~4a), but it is relatively very weak.

As can be seen in Figure~3 in the global spectra the
 peak at $\approx$11-year
 period of SSGA is slightly smaller than that of NSGA, whereas the peak 
of $\approx$50-year period of SSGA slightly larger (close to the 95\% confidence level)
 than that in NSGA (much less than 95\% confidence level). Therefore, 
the $\approx$12.8-year and $\approx$51-year periodicities in RNSA and 
the $\approx$9-year and $\approx$51-year periodicities 
 in ANSA are look  to be resulted due to 
the  aforementioned differences in the strengths of $\approx$11-year 
and $\approx$50-year periodicities in NSGA and SSGA. In addition, 
the approximate two year difference
 exists between the $\approx$12.8-year/$\approx$9-year period of RNSA/ANSA and 
$\approx$11-year period of NSGA and SSGA could be due to 
there exist 1\,--\,2 years differences in the positions  (1\,--\,2-year 
phase difference) of the $\approx$11-year 
peaks of   NSGA and SSGA ($e.g.$, \opencite{ng10}; \opencite{jj19}, also see 
Figure~1). Overall,  the $\approx$12-year and $\approx$51-year
 periodicities
in the north--south asymmetry of solar activity  seem to be manifestations of
the  differences in the strengths of $\approx$11-year
and $\approx$51-year periodicities of activity in northern- and
southern-hemispheres. As already mentioned in Section~1, 
 the statistical error in the relative asymmetry, RNSA, is much 
smaller than  that
of the absolute asymmetry, ANSA (see \opencite{jg97}). Therefore, the 
variations of RNSA can be  more reliable than the variations of ANSA.
 That is,  the $\approx$12.8-year  and $\approx$51-year
 periodicities of RNSA 
are  more reliable than  the $\approx$9-year 
and $\approx$51-year  periodicities of ANSA.  

Figure~5a shows the Morlet wavelet and global
power spectra of   $\overline{\psi_{\rm D}}$ in ten-day intervals during
the periods 1600\,--\,2099. In this figure the epochs of
the  steep decrease in the
orbital angular momentum of the Sun are also shown (also see \opencite{jj05},
\citeyear{jj17}). 
Figure~5b shows  the wavelet and  global
power spectra of   $\overline{\psi_{\rm D}}$ in ten-day intervals during
the periods 1874\,--\,2017. We compare this figure  with Figures~4a and 
4b, $viz.$ the 
wavelet spectra of RNSA and ANSA for the period 1874\,--\,2017.
As can be seen in Figure~5a, there exist $\approx$13.6-year and 
$\approx$41.6-year periodicities in $\overline{\psi_{\rm D}}$ around the years 
1632, 1672, 1811, 1851, 1990, and 2030,  when there was  steep decrease in the 
Sun's orbital angular momentum about the solar system barycenter
caused by some specific  configurations of the giant planets 
(see \opencite{jj05}).  
The well-known 
179-year period of the Sun's orbital angular momentum (\opencite{jose65})
 is obviously exists 
in $\overline{\psi_{\rm D}}$  (however, most of the power spectral region 
 of this period is within the region of  cone of influence).  
In addition, relatively a very weak $\approx$20-year
 periodicity exists in $\overline{\psi_{\rm D}}$,  continuously throughout
 the 1600\,--\,2099.
The wavelet and the global spectra of RNSA and ANSA are somewhat 
closely match with that  of $\overline{\psi_{\rm D}}$ shown Figure~5b.
Particularly, there is a suggestion that the 12\,--\,13-year and 
 40\,--\,50-year
 periodicities in RNSA and ANSA were occurred 
during approximately the same times as the corresponding periodicities
 in $\overline{\psi_{\rm D}}$.
In  the global spectra of both north--south
 asymmetry of sunspot area  and $\overline{\psi_{\rm D}}$ there is a
 suggestion that an increase of 
power with an increase in the value  of period. That is, there is a suggestion 
that  not only 
 the values of the aforementioned periods of the north--south asymmetry of
 solar activity and  $\overline{\psi_{\rm D}}$ and their timings match, the
 relative powers of these periods  are also approximately match.  
 Therefore, we
suggest that there could be influence of some specific  configurations of
the giant planets in  the origin of
 $\approx$12-year and 40\,--\,50-year periodicities of the north--south
asymmetry of solar activity.

 The power spectral area of the $\approx$51 periodicity in RNSA and ANSA  
 is within the region of cone of influence in the corresponding wavelet
 power spectra.  
That is, the available 144 years data of sunspot groups used here are not
 adequate for accurately determining  this 
periodicity in RNSA and ANSA from wavelet analysis. On the other hand,
 as can be seen in Figure~5a, the existence of a 40\,--\,50-year periodicity
 in $\overline{\psi_{\rm D}}$ is very clear in the data during 1600\,--\,2099. 
 In view  of this,  the corresponding periodicity in   
 $\overline{\psi_{\rm D}}$ obtained  from  the  data during the 
period 1874\,--\,2017 
  may be reliable in spite of most of its power spectral area is within
 the region of the cone of influence. This may imply that the existence of 
this periodicity  
seen in the wavelet power spectra of RNSA and ANSA is mostly reliable. 
Moreover, the 50-year peaks in
the Global power spectra  of the north-south asymmetry are statistically
 significant on much more than 95\% confidence level.
This periodicity in north--south asymmetry of solar activity is found to be 
statistically significant in the fast Fourier transform and periodogram 
analyses ($e.g.$, \opencite{jg97}; \opencite{deng16}). 
 The  40\,--\,50-year periodicity in
 $\overline{\psi_{\rm D}}$ is not continuously  present throughout
 the period 1600\,--\,2099 (see Figure~5a). 
 In particular, as already mentioned above,  
this periodicity, and even the $\approx$12-year periodicity,   in 
$\overline{\psi_{\rm D}}$ seem to present
 strongly around the epochs where the Sun's orbital 
motion is
retrograde. Hence, around such occasions the $\approx$12-year and 
 $\approx$51-year  periodicities may be occurring 
 and strong in the north--south asymmetry of solar activity. 

 Figures~6a and 6b show the cross-wavelet power 
  spectrum of RNSA and that of  $\overline{\psi_{\rm D}}$   and
 ANSA and $\overline{\psi_{\rm D}}$, respectively,  during the period 
1874\,--\,2017.
These figures indicate a 
large similarities (covariance)  exist between the time series of  
$\overline{\psi_{\rm D}}$ and north--south asymmetry in sunspot-group area
at scale (period)
$\approx$12-year  mainly before 1940 and  at the scale (period) $\approx$51-year
throughout  1874\,--\,2017. 
Figures~7a and 7b show the wavelet-coherence 
  of RNSA and  $\overline{\psi_{\rm D}}$   and that of 
 ANSA (normalized) and $\overline{\psi_{\rm D}}$, respectively,  during the period
1874\,--\,2017. In these figures there is a suggestion that the
 wavelet-coherence 
exists between  $\overline{\psi_{\rm D}}$ and north--south asymmetry of 
solar activity at most of the known periodicities (including 1.5-year, 
1.8-year, 2.1-year, 3.6-year, $etc.$)  of the latter, at several times.
 However, in   WCOH spectra spurious
peaks can occur for the areas of low wavelet power (\opencite{mk04}). 
On the other hand, the wavelet-cross spectrum may  
not suitable for significance testing the interrelation between
two processes. Therefore, wavelet-coherence may be very useful
(\opencite{mk04}).
Nevertheless,  the wavelet cross spectra (Figures 6a and 6b), and even the 
WCOH spectra (Figures 7a and 7b), suggest that  
in the case of  $\approx$12-year periodicity,   only before 1940 a large
 similarity (covariance)  exists between the 
time series of $\overline{\psi_{\rm D}}$ and north-south asymmetry in 
sunspot-group
 area. In addition,  the  spectral region of $\approx$51-year periodicity 
is within the region of cone of influence. Therefore, the inference above,
 $i.e.$ the influence of some 
specific configurations of the giant planets in the origin of 
 $\approx$12-year and $\approx$51-year  periodicities of 
 north--south asymmetry of  
solar activity  is only suggestive rather than compelling.

\section{Conclusions and Discussion}
The existence of $\approx$12-year and 
$\approx$51-year periodicities
 in the north--south asymmetry of solar activity 
seems to be established to some extent by now from observational 
 point of view. 
 However, the physical reason of these
 as well as the known relatively short 
 periodicities  in  the north--south asymmetry is not yet clear.
Here we have analyzed the combined daily data of sunspot groups reported in  
GPR and DPD during the period 1874\,--\,2017 and the data of 
the orbital positions (ecliptic longitudes) of the giant planets in
 ten-day intervals during the period 1600\,--\,2099. 
Our analysis suggests that the $\approx$12-year and $\approx$51-year
 periodicities
in the north--south asymmetry of solar activity  could be manifestations of 
the  differences in the strengths of $\approx$11-year
and $\approx$51-year periodicities of activity in northern- and 
southern-hemispheres.
During the period 1874\,--\,2017 
the Morlet wavelet power spectra of the north--south asymmetry of 
 sunspot-group area and the
 mean absolute difference  ($\overline{\psi_{\rm D}}$)  of the
orbital positions of the giant planets are found to be similar. 
Particularly, the wavelet spectra 
 suggest that the 12\,--\,13-year and  40\,--\,50-year periodicities 
in the north--south asymmetry of sunspot area   occurred
during approximately the same times as the corresponding periodicities
 in  $\overline{\psi_{\rm D}}$. In addition, the relative powers of 
these periods of the north--south asymmetry and $\overline{\psi_{\rm D}}$ 
are also found to be matched. 
 Therefore, we
suggest that there could be  influence of some specific configurations of 
the giant planets in  the origin of 
the 12\,--\,13-year and 40\,--\,50-year periodicities in the north--south 
asymmetry of solar activity.  

Depending on the torque acting on the system, the angular momentum of the
system might be conserved in only one or two directions but not all
the directions.
The net torque on the Sun (an extended body)  due to gravitational force
of the  planets
may not be always zero, though it is small. That is,  there could
be some changes in the  solar  differential  rotation rate
due to the external forces on the Sun ($e.g.$, \opencite{zaq97}) 
and hence, in the strength of
solar dynamo, suggesting
planetary configurations  have an influence in solar variability.
However, a planetary influence on solar activity, even if it exists,
 seems to be negligibly small~(\opencite{dej05}). 
There is  a continuous progress   on  
this topic.  \inlinecite{abreu12} suggested that  
 variations in solar rotation through 
time-dependent torque exerted by the planets on the non-spherical solar 
 tachocline responsible for the solar cycle and other variations of 
solar activity. \inlinecite{wil13}  constructed 
a Venus--Earth--Jupiter spin--orbit coupling model. According to 
this model  net tangential torques due to time dependent 
alignments of Venus, Earth, and Jupiter    act
upon the outer convective layers of the Sun with periodicities that match 
many of the long-term solar cycles. \inlinecite{stef16} and 
\inlinecite{stef19} have shown that 
tidal force induced by the Venus--Earth--Jupiter system affect solar dynamo 
through  resonant excitation of the oscillation of the $\alpha$-effect.

When the Sun's orbital angular momentum
was close to zero (the Sun was close to the barycenter), the solar
 equatorial rotation rate was also considerably low~(the steep
decreases in   the Sun's spin and orbital angular momenta  have
the same value,
$\approx$$10^{47}$ g cm$^2$ s$^{-1}$; see \opencite{jj05}), suggesting that the
differential rotation rate was  considerably low, $i.e.$ a weak dynamo.
That is, at least
the  Sun's retrograde orbital motion about the solar system barycenter
may  influence the solar dynamo and  may be  responsible
for the grand minima, such as Maunder minimum (also see \opencite{cc12}).
 Although the altitude of planetary tidal waves on the
Sun is of the order of only
one millimeter~($e.g.$, \opencite{cs75}),   it may be enlarged appreciably 
at the times of 
 some  alignments of the Jupiter with other planets.
There may be a complicated coupling between the variation in the rotation
rate caused by  some specific alignments of the tide rising planets
(Venus, Earth, and Jupiter) and the variation in
the orbital motion of the Sun due to  some specific alignments of the
giant planets.
 Depending  on the differences
among the angular inclinations of the orbits of planets to the ecliptic
(in the heliocentric co-ordinate system), during a specific 
 configuration of planets
 the maximum angular distances of the planets above and below the
ecliptic (or invariant plane) cause  
the maximum north--south asymmetry in the distribution of mass and
angular momentum in the solar system. 
\inlinecite{jj03} showed the presence of the  periods of the 
alignments of two or more  giant planets in both
 the solar differential rotation and  its north--south asymmetry 
 determined from 
the sunspot-group data during the period 1879\,--\,1976.
   As suggested by 
\inlinecite{gj95},   variations  in the 
north--south asymmetry of solar activity may represent  anti-symmetric 
 global modes of solar magnetic oscillations and the perturbations 
 needed for all the global modes of solar magnetic oscillations may be 
 provided by some specific
configurations of the  planets. Moreover, besides there exists a considerable 
north--south asymmetry in both the solar differential rotation and the 
 meridional flow, 
 a reasonable correlation also exists between 
 solar cycle variations of the differential rotation rate and 
the meridional motion of sunspot groups~(\opencite{ju06}).
\inlinecite{jhs17} found the existence of  
 a highly statistical significant  correlation  between the
solar  meridional flow and  the Sun's orbital
torque during the Solar Cycle~23.
Recent numerical simulations from a flux transport dynamo model
 show importance of
time variation and north--south asymmetry in meridional circulation
 in producing differing solar cycles in the northern- and southern-hemispheres
 (\opencite{bd13}).  
 We think that  
the effect of differences in the  time-dependent configurations of  planets
 above and below the 
ecliptic may be responsible for 
variations in the north--south asymmetry of 
solar differential rotation and meridional circulation that are in turn
 responsible for   the variations of
north--south asymmetry in solar activity.

\section{Acknowledgments}
The author thanks  anonymous referee for helpful comments and  suggestions.
The author is thankful to Ferenc Varadi for providing
the entire planetary data used here. Wavelet software was provided
 by C. Torrence and G. Compo
 and is available at {\tt http://paos.colorado.edu/research/wavelets}.

\section{Disclosure of Potential Conflicts of Interest}
The author declares that he has no conflicts of interest.

{}
\end{article}
\end{document}